\documentclass[pre,twocolumn,groupedaddress]{revtex4}
\usepackage{graphicx}
\begin{document}
\title{Thermodynamics of Community Structure} 
\author{Claire P.~Massen}
\affiliation{Department of Chemistry, University of Cambridge, 
Lensfield Road, Cambridge CB2 1EW, United Kingdom}
\author{Jonathan P.~K.~Doye}
\email{jonathan.doye@chem.ox.ac.uk}
\affiliation{Physical and Theoretical Chemistry Laboratory,
Oxford University, South Parks Road, Oxford OX1 3QZ, United Kingdom}
\date{\today}

\begin{abstract}

We introduce an approach to partitioning networks into communities that not 
only determines the best community structure, but also provides a range of 
characterization techniques to assess how significant that structure is.
We study the thermodynamics of community structure by producing equilibrium ensembles of partitions, in which each partition is represented with a well-defined statistical weight.
Thus we are able to study the temperature dependence of thermodynamic properties, namely the modularity $Q$ and heat capacity, with particular emphasis on the transition between high-temperature, essentially random partitions and low-temperature partitions with high modularity.
We also look at frequency matrices that measure the likelihood that two nodes belong to the same community, and introduce an order parameter to measure the `blockiness' of the frequency matrix, and therefore the uniqueness of the community structure. 
These methods have been applied to a number of model networks in order to understand the effects of the degree distribution, spatial embedding and randomization.
Finally, we apply these methods to a metabolic network known to have strong community structure and find hierarchical community structure, with some communities being more robust than others.

\end{abstract}

\maketitle

\section{Introduction}
\label{sec:intro}

Many networks do not have a homogeneous topology, but are composed of interacting modules of nodes with a high density of edges within the modules.
For example, in the world wide web, web sites covering similar topics group together into modules \cite{Flake02}.
In social networks, communities naturally form, where people having certain characteristics in common are more likely to be acquainted.
Examples include communities of scientists working on similar areas of research \cite{Girvan02, Newman03c}, jazz musicians grouped by race \cite{Gleiser03}, departments in an organization \cite{Guimera02} and authors of home pages who have some common interests \cite{Adamic03}.
Many non-social networks also have underlying community structure.
For example, biological networks often consist of functional modules \cite{Hartwell99,Barabasi04,Ma04,Guimera05}.

These modules can be determined purely from the topology of the network, removing human bias.
Knowledge of the modules can then lead to further understanding of a network.
For example, nodes in the same community usually have some properties in common, facilitating classification of nodes.
This can be used to find similar web pages \cite{Flake02}, aiding search engines and content filtering.
In biological networks, it can be used to find metabolites, proteins or genes performing similar functions, and to understand the role of different elements \cite{Hartwell99,Milo02,Wilkinson04,Barabasi04,Ma04,Guimera05,Palla05,Huss06}.

Processes occurring on a network are affected by its modularity, for example communication patterns \cite{Granovetter73,Tyler03book} or web searching \cite{Xie04}.
Trends spread well within communities \cite{Bettencourt03}, but diffusion between different communities is slow \cite{Eriksen03, Latapy04, Simonsen04}.
The amount of time an epidemic lasts on a network has been shown to be a function of the number of modules \cite{Huang06}.
Furthermore, although many networks have short paths, including random networks, the short paths can be very difficult to find.
Communities in a social network assist in finding short paths \cite{Watts02,Motter03}.
Modularity can be also used to simplify the amount of data in large complex 
networks,
for example, through coarse-graining \cite{Newman03c, Guimera05, Song05, Itzkovitz05} or through characterization of networks by studying the occurrence of motifs \cite{Milo02}.

For networks with strong community structure, a key question is how significant or real is that community structure.
Most studies on community structure involve finding the `best' partition of nodes into communities \cite{Newman04, Danon05}, where best is usually defined in terms of a quantity called the modularity \cite{Newman03c,Newman03d}.
Occasionally comparison to random networks with the same degree distribution is made to try to assess the significance of any observed modularity \cite{Guimera04, Massen05, Huss06}.
However, by studying just the one best partition, with each node assigned to one community, a lot of useful information is lost.
Recently it has become clear that many networks are not simply composed of distinct, weakly interacting modules \cite{Reichardt06b}.
For example, the modules could be `fuzzy', in the sense that some nodes belong to more than one community, as in social networks where people belong to many overlapping communities \cite{Wilkinson02,Tyler03book,Reichardt04,Palla05,Derenyi05,Gfeller05}.
The modules could also join together, forming larger modules in a hierarchical manner \cite{Ravasz02,Guimera02,Holme03,Ravasz03,Zhou03,Muff05}.

A further consideration is the robustness or uniqueness of a given partition.
Robustness looks for not just the best community structure, but whether it is well defined.
In some cases, there can be many partitions with similarly high modularity, but with a very different partitioning of nodes into communities.
This indicates that the community structure is probably not particularly informative, an extreme example being lattices.
The networks based on lattices are of course very uniform, with identical environments around each lattice point.
However, applying a community detection algorithm to a lattice uncovers partitions with high modularity \cite{Guimera04, Massen05}.
If just judged in comparison to a randomized graph, it would have been concluded that this community structure is significant.
However, any partition of a lattice is non-unique, since translation of the community boundaries by any lattice vector will generate a new partition with identical modularity.
Although the lack of uniqueness in the example of a lattice is transparent, it would be useful to have a general method to test for the lack of robustness in obtained community structures.

In this work, we study equilibrium ensembles of communities at different temperatures.
Previous works have begun to look at ensembles of community structures \cite{Wilkinson02,Girvan02,Newman03c,Duch05,Guimera05,Reichardt06}.
However, the ensembles have usually been generated in an {\it ad hoc} manner, for example by introducing a stochastic element into the algorithms or by comparing the partitions obtained at the end of a number of runs of the algorithm.
Here, the ensembles are sampled in a statistically rigorous way, with each partition in the ensemble having a well-defined weight.

The use of temperature is also implicit to some methods for optimizing community structure, with better partitions found at lower temperatures \cite{Guimera04,Reichardt04,Guimera05,Massen05,Medus05,Muff05}.
Here, we rephrase the question of how unique is the obtained community structure as how different are the partitions obtained at a given temperature.
These differences can be visualised using a matrix showing how many times each pair of nodes are classified as being in the same community \cite{Reichardt04,Guimera05, Duch05}.
Uniqueness, hierarchical and fuzzy community structures can be seen clearly.

In Section \ref{sec:methods}, we first introduce the methods used and the properties we measure.
Then in Section \ref{sec:test}, we study computer-generated test networks with prescribed community structure, introduced by Girvan and Newman and often used to test algorithms \cite{Girvan02,Newman03c,Danon05}.
To illustrate the issue of uniqueness of community structure, a hexagonal lattice is studied in Section \ref{sec:lat}.
We then apply our approach to scale-free networks in order to determine the effect of the degree distribution.
The Apollonian network (Section \ref{sec:apollo}) is an example of a spatial, scale-free network \cite{Andrade05,Doye05}.
It has some community structure \cite{Doye05}, grouping together nodes that are close in space as for the hexagonal lattice, but with different length scales potentially leading to a hierarchical structure.
In Section \ref{sec:sf}, we study scale-free test networks with introduced community structure similar to the original test networks, the difference being that the original test networks have an Erd\"os-Renyi degree distribution.
Finally, in Section \ref{sec:met} we apply the algorithm to a metabolic network describing the interactions between metabolites in \emph{E.\ coli} \cite{Ma03}.
Community structure has been uncovered in this network previously \cite{Ma04,Ma04b,Guimera05,Muff05}.
However, biological networks often have complex, hierarchical community structure rather than isolated modules \cite{Ravasz02,Holme03b,Barabasi04}.
As such, our approach is particularly appropriate.

\section{Methods}
\label{sec:methods}

To find good partitions the modularity is often used \cite{Newman03c,Newman03d}, and is defined by
\begin{equation}
Q=\sum_c \left( e_c/M - \left( \sum_{i \in c} k_i/2M \right)^2 \right).
\label{eqtn:q}
\end{equation}
\noindent This quantity gives a measure of how many more edges are within communities compared to that expected for a random network.
$M$ is the total number of edges in the network, and $k_i$ is the degree
of node $i$.
$e_c$ measures the number of edges lying within community $c$, i.e.~the nodes at either end of these edges are both classified as belonging to community $c$ for this partition.
$\sum_{i \in c} k_i/2M$ is the fraction of all ends of edges that arrive at community $c$.
If an edge is chosen and followed at random, it gives the probability that the node at the end belongs to community $c$.
Hence, $\left( \sum_{i \in c} k_i/2M \right)^2$ gives the predicted fraction of edges within community $c$.

To characterize the possible partitioning of the network into communities, we generate a canonical ensemble of network partitions, with $-Q$ playing the role of energy.
i.e.~at temperature $T$, the statistical weight of a given partition in the ensemble is proportional to $\exp(Q/T)$.
Each partition is described by a vector assigning each node to a community, similar to the coordinate vector describing a configuration.
The ensemble is simulated using Metropolis Monte Carlo.
Each move involves changing the community assigned to a node.
If the move leads to a better community structure, i.e.~an increase in $Q$, the move is accepted, otherwise it is accepted with a temperature-dependent probability $\exp(\Delta Q /T)$, where $\Delta Q $ is the difference in $Q$ between the new and old partitions.

For a network with $N$ nodes and $n_c$ communities, a node is chosen with probability $1/N$ and assigned to either one of the existing communities or to a new community, each with probability $1/(n_c+1)$.
Because the aim is to generate equilibrium ensembles of partitions, detailed balance must be maintained, i.e.~the probability of attempting a move (not necessarily accepting it) must be equal to the probability of attempting the reverse move.
If the selected node is in a community of size one, i.e.~by itself, a move will either leave the structure unchanged, or decrease the number of communities.
Therefore, a community must be chosen with probability $1/n_c$, excluding the option of starting a new community, which in this case is equivalent to choosing the community to which the node already belongs.

To facilitate the generation of equilibrium ensembles even at low temperature, we use parallel tempering \cite{Earl05}.
In this method, Monte Carlo simulations are simultaneously run at a set of different temperatures, but where additional exchange moves are attempted that involve swapping the partitions between two temperatures.
Such moves are always accepted if the higher temperature had a higher $Q$ value, otherwise it is accepted with probability $\exp(-\Delta \beta \Delta Q)$, where $\Delta Q$ is now the difference in $Q$ between the partitions at 
the two temperatures and $\Delta \beta$ is the difference in the inverse temperature $\beta$.

The initial state of each ensemble has each node in a separate community.
A number of steps are then required to equilibrate the system.
At the lowest temperature, $Q$ tends to a single, maximum value, $Q_{max}$.
We call the corresponding partition the global maximum.
Although without testing all of the possible partitions it is impossible to be certain there is not a partition with higher $Q$ \cite{Brandes06}, the use of the parallel tempering method makes it likely that the true global maximum has been found for the network sizes used here.

Various properties can be calculated for these ensembles.
We study the variation of $Q$ with temperature, seeing an increase towards an optimal value at low temperatures.
The heat capacity, $C= -dQ / dT$ is used to characterize the nature of these changes with a peak in the heat capacity representing a transition.
The more rapidly that $Q$ changes, the sharper the peak will be.
We define the transition temperature, $T_c$, as the temperature at which this peak occurs.

We also introduce an order parameter to measure the similarity of the sampled partitions at a given temperature, reflecting the uniqueness of the underlying community structure.
We look at frequency matrices, measuring the relative number of times each pair of nodes is classified as belonging to the same community.
If only one community structure is observed, as is generally the case at sufficiently low temperatures, the same pairs of nodes are assigned to the same community in each partition, giving a block diagonal matrix.
At infinite temperatures, each partition has equivalent weight and any pair of nodes is equally likely to be together, giving a homogeneous matrix.
At intermediate temperatures, there may be some remnants of community structure, with some pairs of nodes being assigned to the same community relatively often, showing as darker areas in the frequency matrix.

This uniqueness can be quantified using the Fiedler eigenvalue $\lambda_2$, \cite{Fiedler73,Newman04} i.e.~the magnitude of the second smallest eigenvalue (the smallest being zero) of the Laplacian.
The elements of the Laplacian matrix are the negative of the elements of the frequency matrix, except for the diagonals, which are sums over the corresponding row.
For comparisons between different temperatures, the frequency matrix is first rescaled by the sum of all the elements, which is larger when there are fewer, larger communities at lower temperatures.
$\lambda_2$ gives a measure of how `diagonal' the matrix is, and takes values of zero for a block diagonal matrix and one for a homogeneous matrix.
It therefore gives a measure of how different the partitions at each temperature are and can be used as an order parameter, since it is close to one at high temperatures and decreases at low temperatures.
If the best community structure, seen at low temperatures, is degenerate, i.e.~there are competing partitions with equal values of $Q$, then $\lambda_2$ will remain above zero as the temperature decreases to zero.
Even if $\lambda_2$ does go to zero, it provides insight into the character of the transition.
A sharp transition indicates that there are few competing structures with high $Q$, whereas a broader transition implies the opposite.

The eigenvector corresponding to the Fiedler eigenvalue gives the best order in which to place the nodes to make the frequency matrix look maximally `blocky', and assists with assigning the nodes to communities.
Interestingly, the Fiedler eigenvalue is also used in spectral partitioning to find communities \cite{Newman04,Capocci04}, the difference being that spectral partitioning sorts the adjacency matrix into its most blocky form, whereas here we apply it to the frequency matrix.
It has also been recently used to find communities by maximising $Q$ \cite{Newman06,Newman06b}.

\section{Test networks}
\label{sec:test}

The test networks used consist of 128 nodes, each designated as belonging to one of four communities of 32 nodes each \cite{Newman03c}.
1024 edges are added.
Of these, $1024\,P_{in}$ connect nodes belonging to the same community and the remaining $1024 (1-P_{in})$ connect nodes in different communities.
$P_{in}$ is varied between 0.25 and 1.
When $P_{in}=1$, nodes are only connected to other nodes in the same community, giving four components corresponding to the four communities.
When $P_{in}=0.25$, one quarter of the edges leading from a node are expected to be connected to nodes in the same community.
Based on the sizes of the communities, this is exactly as would be expected if the nodes were connected at random.
Therefore, when $P_{in}=0.25$ the network is essentially random, i.e.~an Erd\"os-Renyi graph.

\begin{figure}
\centerline{\includegraphics[width=8.6cm]{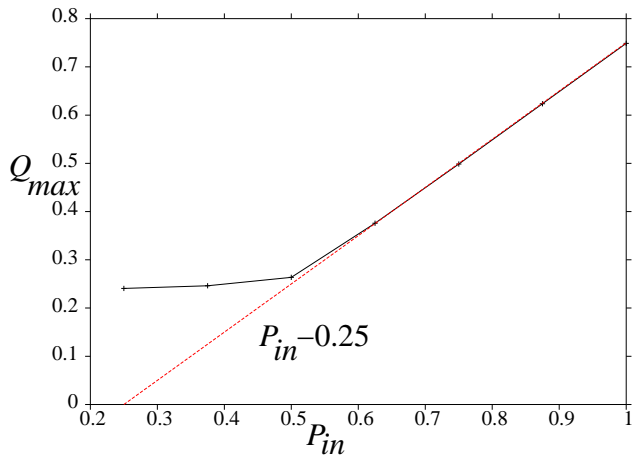}}
\caption{(Colour online)
The highest observed value of $Q$, $Q_{max}$, against $P_{in}$ for the test networks.
For $P_{in} > 0.5$, $Q_{max}$ increases with $P_{in}$ as $P_{in}-0.25$ (also shown).
}
\label{fig:qgm}
\end{figure}

$P_{in}=0.5$ is an interesting case in that a node is as likely to be connected to nodes in its own community as it is to nodes in other communities.
Indeed, we do see some differences between $P_{in} \le 0.5$ and $P_{in}>0.5$.
Assuming that all nodes have the average degree, it can then be shown that the modularity for the input community structure would be given by $Q_{input}=P_{in}-0.25$.
Fig.~\ref{fig:qgm} shows that $Q_{max}$ follows this expression for $P_{in} > 0.5$.
However, for $P_{in} \le 0.5$, we find a global maximum (GM) that is better (has higher $Q$) than the input community structure.
This is because random networks often have partitions with reasonably high modularity as fluctuations in the link distributions are bound to give some partitions with $Q$ significantly above 0 \cite{Guimera04}.
The value of $Q_{max}$ for $P_{in} \le 0.5$ has been predicted to be $Q_{max} \approx 0.21$ by Guimera \emph{et al.}\ \cite{Guimera04} and $Q_{max} \approx0.23$ by Reichardt and Bornholdt \cite{Reichardt06b}.
Both are in line with the result obtained here ($Q_{max} \approx 0.24$), which is for just one realization of the possible networks that have $P_{in}=0.25$.

\begin{figure}
\centerline{\includegraphics[width=7.7cm]{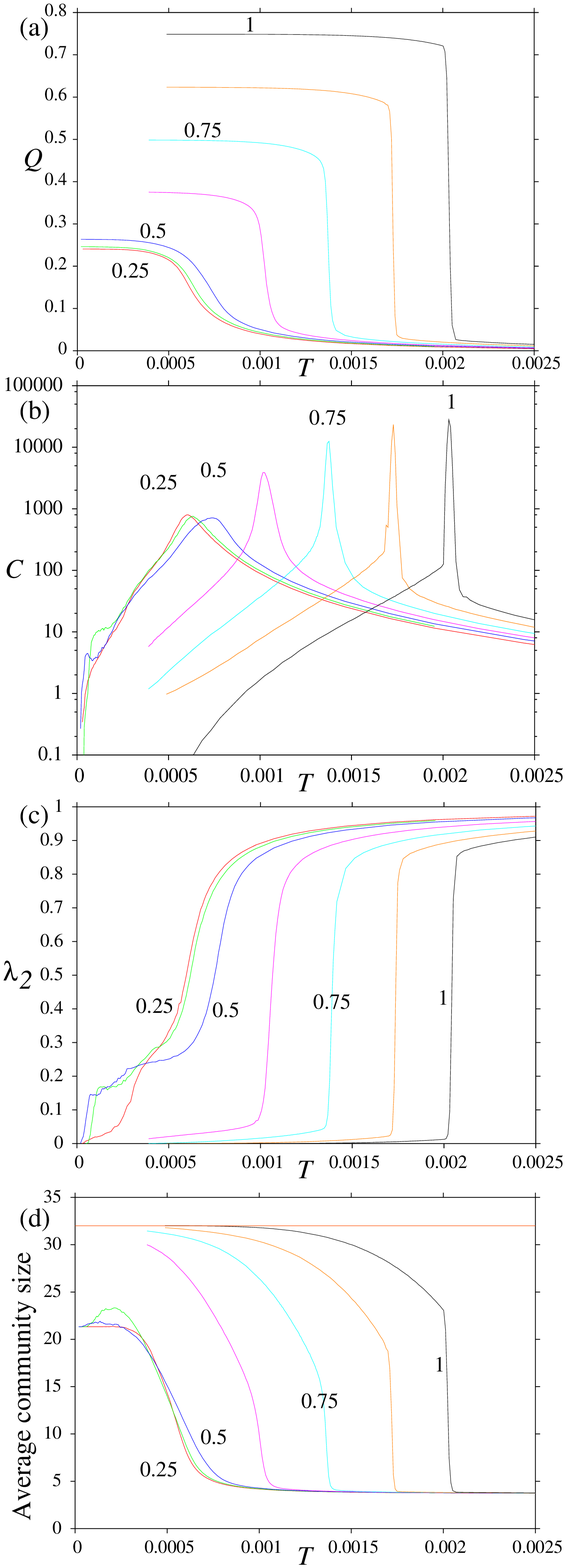}}
\caption{(Colour online) 
Thermodynamic properties for test networks.
Variation of (a) $Q$, (b) heat capacity $C$, (c) $\lambda_2$ and (d) average community size with temperature.
In (d), the community size of the input community structure, which is 32, is also shown.
Equally spaced values of $P_{in}$ have been used, with every other line labelled by its $P_{in}$ value.
Note that in (b) the heat capacity has been plotted on a log scale, unlike in the following figures.
}
\label{fig:test}
\end{figure}

For these networks, there is a clear transition as a function of temperature between a high $Q$, low entropy phase at low temperature, and a low $Q$, high entropy phase at high temperature, as can be seen in Fig.~\ref{fig:test}.
This transition to stronger community structure at low temperature, which is indicated by a peak in the heat capacity in Fig.~\ref{fig:test}(b), is sharper for high $P_{in}$, and becomes similar to a first-order phase transition.
This feature implies cooperativity, i.e.~on heating, different communities break up at the same temperature.
This is because the community structure in these test networks is very homogeneous: all nodes have roughly the same degree, and the input communities all have the same size and strength.

By contrast, for $P_{in} \le 0.5$, the GMs consist of different sized communities, and because of these heterogeneities the transition is much broader, with some communities breaking up before others.
As the GM, $Q_{max}$ and the number of edges within communities is similar for each of these networks, the position and form of the peak becomes approximately independent of $P_{in}$ for $P_{in} \le 0.5$.

For $P_{in} > 0.5$, the transition occurs at higher temperatures for higher $P_{in}$ as the communities are more difficult to break up on heating, and easier to find on cooling.
The `energy cost' (decrease in $Q$) to break up a community depends on how cohesive that community is, and by design is larger for higher $P_{in}$, leading to a higher transition temperature.

\begin{figure}
\centerline{\includegraphics[width=8.6cm]{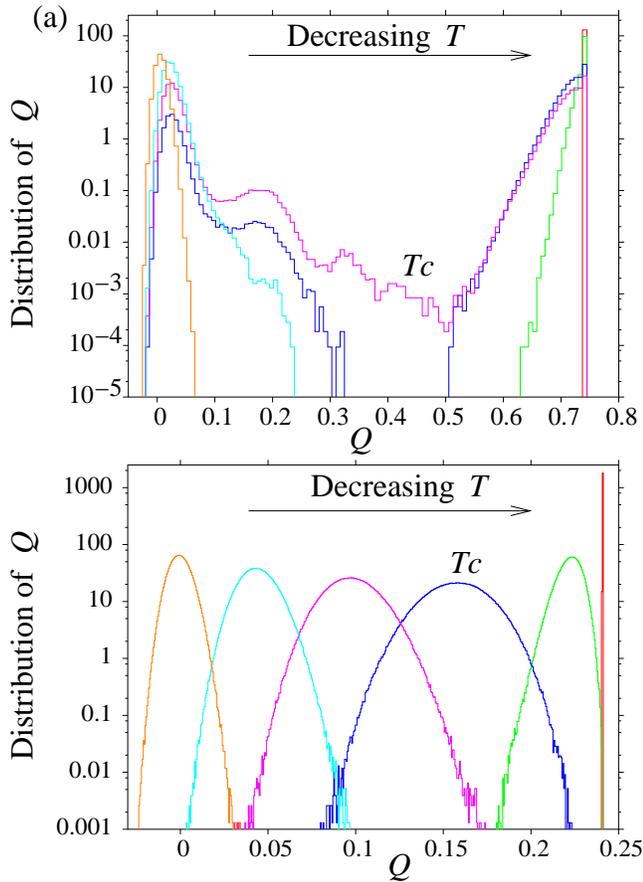}}
\caption{(Colour online) 
Distributions of $Q$ seen at different temperatures for test networks with
(a) $P_{in}=1$ and
(b) $P_{in}=0.25$.
$T_c$ is the transition temperature indicated by a peak in the heat capacity.
}
\label{fig:nij}
\end{figure}

Distributions of $Q$ at different temperatures are shown for $P_{in}=0.25$ and $1$ in Fig.~\ref{fig:nij}.
At high $P_{in}$ the distributions in the transition region are bimodal, 
which is a characteristic of a first-order phase transition.
As few partitions are observed with intermediate values of $Q$, the system switches rapidly between high and low $Q$ partitions,
leading to a sharper transition. 
By contrast, at lower $P_{in}$ the distributions are always unimodal,
and the modal value smoothly changes across the transition.

Possible alternative community structures to the GM would involve either some nodes swapping between different communities or groups of nodes breaking away from the main communities.
Both of these are harder for higher $P_{in}$, where there are more links to the main community and fewer links to nodes from other communities.
Therefore, at $P_{in}=1$ the only alternative seen with any statistical significance are thermal fluctuations about the GM.
As $P_{in}$ decreases, the preferred community of a node becomes less well-defined as it has more connections to other communities.
Consequently, some frustration is introduced, thus making alternative partitions more reasonable, i.e.~they have higher modularity and are sampled at intermediate temperatures.
This leads to the transition becoming broader as $P_{in}$ decreases.

These changes are also reflected in the temperature dependence of the average community size (Fig.~\ref{fig:test}(d)).
For $P_{in}=1$, at $T_c$ there is a sharp jump to larger communities, although only to approximately 23 nodes per community rather than 32 as in the GM.
A typical partition at $T_c$ consists of four large communities similar to those in the GM, with a few very small `communities' of one or two nodes alongside.
There is a heavily preferred scale to the community structure, and very different sizes are hardly ever seen.
As the temperature decreases, these fluctuations die away and the partitions settle to the GM, where all communities have size 32.

\begin{figure}
\centerline{\includegraphics[width=8.6cm]{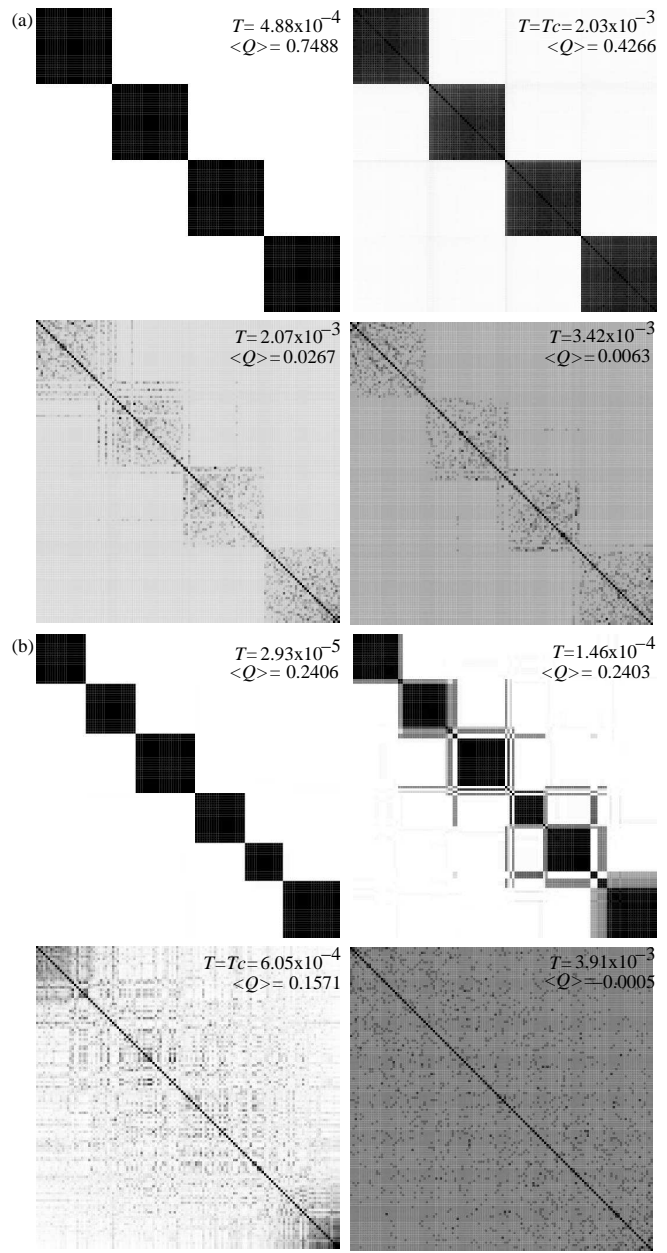}}
\caption{
Frequency matrices at four different temperatures for test networks with (a) $P_{in}=1$, (b) $P_{in}=0.25$. 
}
\label{fig:fmat}
\end{figure}

For lower $P_{in}$, the jump at $T_c$ becomes less pronounced, and immediately below $T_c$ the average community size is smaller.
The GM is less dominant because nodes have more edges to nodes outside their prescribed community, so they are less strongly attached to that community.
Rather than fluctuations about the GM just occurring, communities with a range of sizes are observed.

Frequency matrices for ensembles at selected temperatures are shown in Fig.~\ref{fig:fmat}.
At the lowest temperature, the matrices are block diagonal, because all partitions in the ensemble have the same community structure.
Each block corresponds to one community.
As the temperature is increased, the ensembles consist of partitions with different community structure and the matrices become more homogeneous, with some clustering around the diagonal.
At the highest temperature, the partitions are essentially random and all pairs of nodes are equally likely to be together, giving a very homogeneous matrix.

For $P_{in}>0.5$ the GM is the four input communities of equal size.
On increasing the temperature the partitions remain similar up to $T_c$.
Traces of the underlying community structure are even clear at high temperatures when only small communities are present, reflecting the preference for nodes to mix with nodes from the same predefined communities.
For $P_{in} \le 0.5$ the GM has six communities of different sizes.
For $P_{in}=0.25$, some nodes show only a weak preference for their own community, and at low temperature can switch between two or three different communities, giving rise to the strips evident in Fig.~\ref{fig:fmat}(b), which are a signature of overlapping community structure.
There are also some small communities that have broken away from the main communities.
Although at $T_c$ $Q$ has only decreased by 35\% from its maximum value, the original community structure has virtually disappeared to be replaced by an ensemble of very different partitions, thus giving a fairly homogeneous frequency matrix.

\begin{figure}
\centerline{\includegraphics[width=8.6cm]{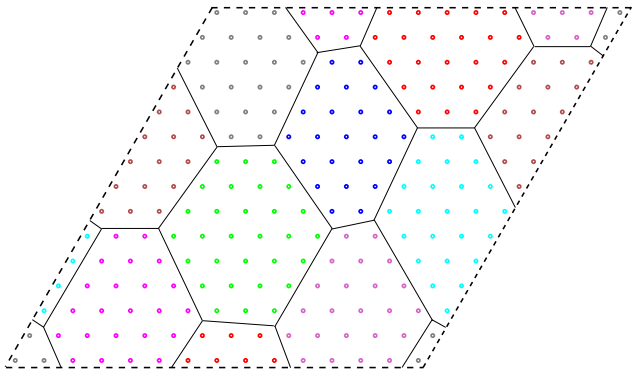}}
\caption{(Colour online) 
Best community structure found for the hexagonal lattice, having $Q=0.6520$. 
Each node has degree 6.
Periodic boundary conditions are used, and a single cell containing 225 nodes is illustrated.
The solid lines represent boundaries between communities, and the dashed lines the boundaries of the cell.
}
\label{fig:latgm}
\end{figure}

The variation from homogeneous to block diagonal in the frequency matrices is reflected in $\lambda_2$, which is close to one at high temperatures and approaches zero at low temperatures, as shown in Fig.~\ref{fig:test}(c).
The GM is unique for each of these networks, so $\lambda_2$ must be zero at $T=0$.
However, the approach to zero differs.
For $P_{in}=1$, the transition is very sharp and $\lambda_2$ is close to zero immediately below the transition, reflecting the uniqueness and strength of the community structure, and the smallness of the fluctuations about the input communities.
For $P_{in} \le 0.5$, $\lambda_2$ is significantly above zero for temperatures well below $T_c$, because in these more random graphs there are different partitions with similarly high values of $Q$.
Thus, $\lambda_2$ is able to detect this non-uniqueness and lack of robustness in the communities in an effectively random graph.

\section{Regular lattice}
\label{sec:lat}

We study a regular lattice next because it shows as a false positive in community detection algorithms, i.e.~there are partitions with high values of $Q$ but the network has no real community structure \cite{Guimera04,Massen05}.
We apply the method to a regular hexagonal lattice with 225 nodes, each with degree $k=6$, using periodic boundary conditions (Fig.~\ref{fig:latgm}).

\begin{figure}
\centerline{\includegraphics[width=8.6cm]{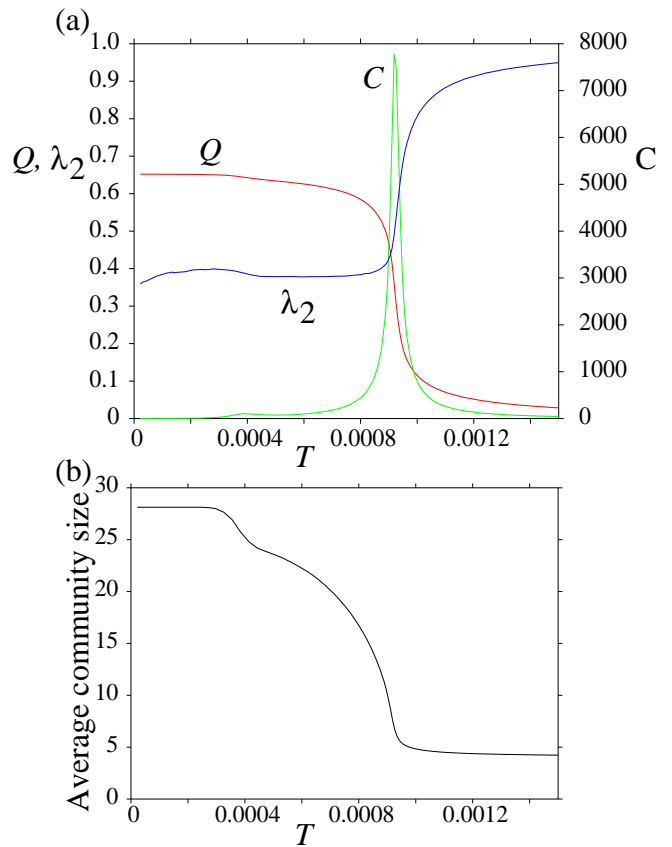}}
\caption{(Colour online) 
Variation of
(a) $Q$,
heat capacity $C$, and 
$\lambda_2$ 
and (b) average community size
with temperature for the hexagonal lattice.
}
\label{fig:lat}
\end{figure}

\begin{figure}
\centerline{\includegraphics[width=8.6cm]{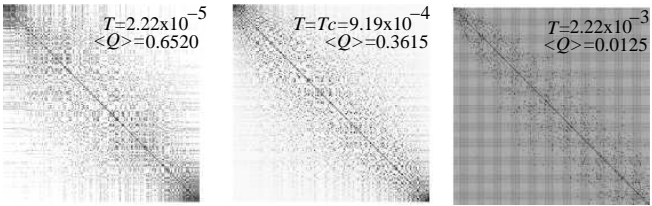}}
\caption{
Frequency matrices at different temperatures for the hexagonal lattice. 
}
\label{fig:fmat-lat}
\end{figure}

The GM that we obtain has high $Q$, and is shown in Fig.~\ref{fig:latgm}.
In the GM, the network is partitioned into hexagons that are close to regular, because this structure minimizes the number of edges between communities.
As the GM is degenerate---translating the boundaries of these communities gives a new partition with the same high value of $Q$---$\lambda_2$ does not tend towards zero at low temperature (Fig.~\ref{fig:lat}).
Similarly, the frequency matrix at low temperature is not block diagonal because of the many different high $Q$ partitions (Fig.~\ref{fig:fmat-lat}), but neither is it entirely homogeneous, because nodes that are close together in space are more likely to be assigned to the same community.

\begin{figure}
\centerline{\includegraphics[width=8.6cm]{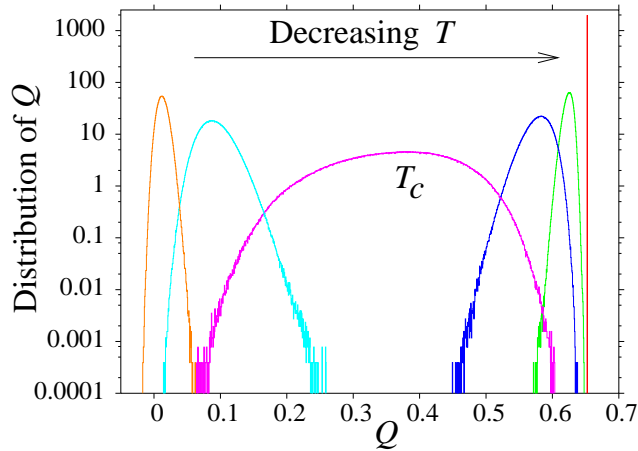}}
\caption{(Colour online)
Distributions of $Q$ at different temperatures for the hexagonal lattice. 
}
\label{fig:latnij}
\end{figure}

Despite these features, the transition is relatively sharp, and  $Q$ remains large until close to the transition.
Unlike the model networks with higher $P_{in}$ considered in the last section, partitions with intermediate values of $Q$ can easily be generated by decreasing the size of the spatially-localized communities.
Reflecting this difference, the distribution of $Q$ is unimodal at all temperatures (Fig.~\ref{fig:latnij}).
At $T_c$, the distribution is very broad, sampling lots of partitions with a wide range of $Q$.

Similarly, there are no jumps in the temperature dependence of the average community size as in Fig.~\ref{fig:test}(d), but instead it decreases relatively smoothly with temperature, with the two heat capacity peaks corresponding to the most rapid decreases in the community size (Fig.~\ref{fig:lat}).
The largest heat capacity peak is sharp, probably reflecting the uniformity of the network, which makes it more likely that changes to the community structure occur simultaneously throughout the network.

\section{Apollonian packing}
\label{sec:apollo}

\begin{figure}
\centerline{\includegraphics[width=8.6cm]{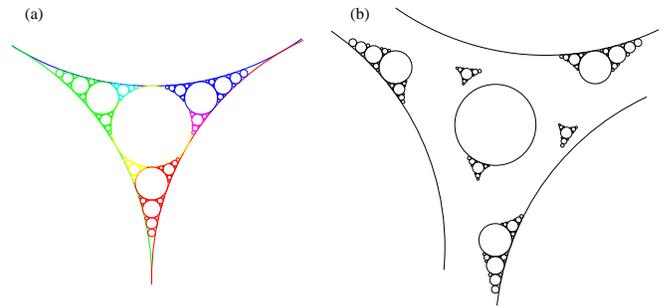}}
\caption{(Colour online) 
Best community structure found for an Apollonian network, having $Q=0.5030$.
The initial configuration for generating the Apollonian packing is three disks all touching each other.
Then in the interstice between them, the largest possible disk is inserted, creating three new interstices.
These interstices can in turn be filled.
This process can be iterated until the whole of space is completely filled by disks.
In the Apollonian network the disks correspond to the nodes, and edges occur wherever two disks are in contact.
In the packing illustrated, disk insertion was stopped after five iterations, giving a finite network with 124 nodes.
The communities are represented in (a) by different shades and in (b) the communities have been separated for clarity.
}
\label{fig:apgm}
\end{figure}

The network of contacts between disks in an Apollonian packing is an example of a scale-free, hierarchical and spatial network \cite{Andrade05,Doye05}.
The packing is a fractal object \cite{Mandelbrot}, and space is filled with different-sized disks, as shown in Fig.~\ref{fig:apgm}.
Similar to the regular lattice, this Apollonian network has community structure with high $Q$ (Fig.~\ref{fig:apgm} shows the GM) due to its spatial nature, and the resulting communities are spatially localized. 
However, $Q_{max}$ is not as large as for the regular lattice of the previous section, probably because scale-free networks are inherently less separable because high-degree nodes connect up many parts of the network.

\begin{figure}
\centerline{\includegraphics[width=8.6cm]{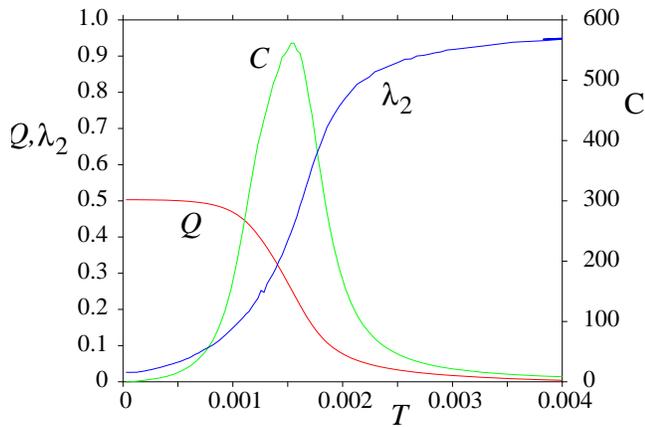}}
\caption{(Colour online) 
Variation of
$Q$,
heat capacity $C$, and 
$\lambda_2$ 
with temperature for the Apollonian network.
}
\label{fig:ap}
\end{figure}

Fig.~\ref{fig:ap} shows that the thermally-induced transition is much broader and less cooperative for the Apollonian network than for the test networks and the regular lattice.
In those networks the degree distribution is very narrow, each node has a similar degree, and the communities also have similar sizes.
On heating, each node is similarly likely to break away from its community at the same temperature, leading to a sharp transition.
In a scale-free network, where the nodes have very different degrees, a low-degree node may be more likely to break away from its community and start a new one on heating, whereas a high-degree node is likely to have more connections to its community and so starting a new community would involve a large decrease in $Q$.
On the other hand, a high-degree node may be more likely to switch to a different community, as it probably also has a significant number of connections to other communities, whereas a low-degree node could easily have all of its connections within one community and so strongly prefer that community.

\begin{figure}
\centerline{\includegraphics[width=8.6cm]{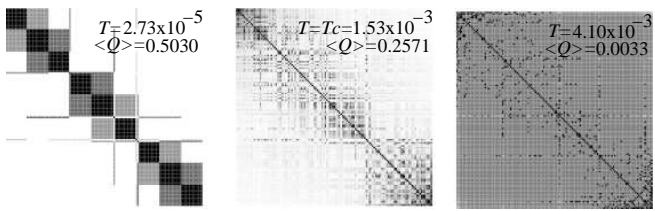}}
\caption{
Frequency matrices at different temperatures for the Apollonian packing.
}
\label{fig:fmat-ap}
\end{figure}

$\lambda_2$ (Fig.~\ref{fig:ap}) is much closer to zero at low temperatures than for the lattice, reflecting the more block diagonal frequency matrix (Fig.~\ref{fig:fmat-ap}), but it does not not reach zero, because there is some degeneracy due to the three-fold symmetry of the packing.
The low-temperature frequency matrix reflects the hierarchical nature of the modularity, i.e.~there are competing `length scales' for the community structure.
The chequerboard pattern on the diagonal is due to competing partitions that have only marginally different $Q$ values, the difference being whether to group or break up certain communities.
There are some thin strips of colour corresponding to nodes that can belong to any of a range of communities, providing some overlap between them.
These are the hubs in the network, namely the large central disk and the three large peripheral disks (only partially shown in Fig.~\ref{fig:apgm}).

\section{Scale-free test networks}
\label{sec:sf}

\begin{figure}
\centerline{\includegraphics[width=8.6cm]{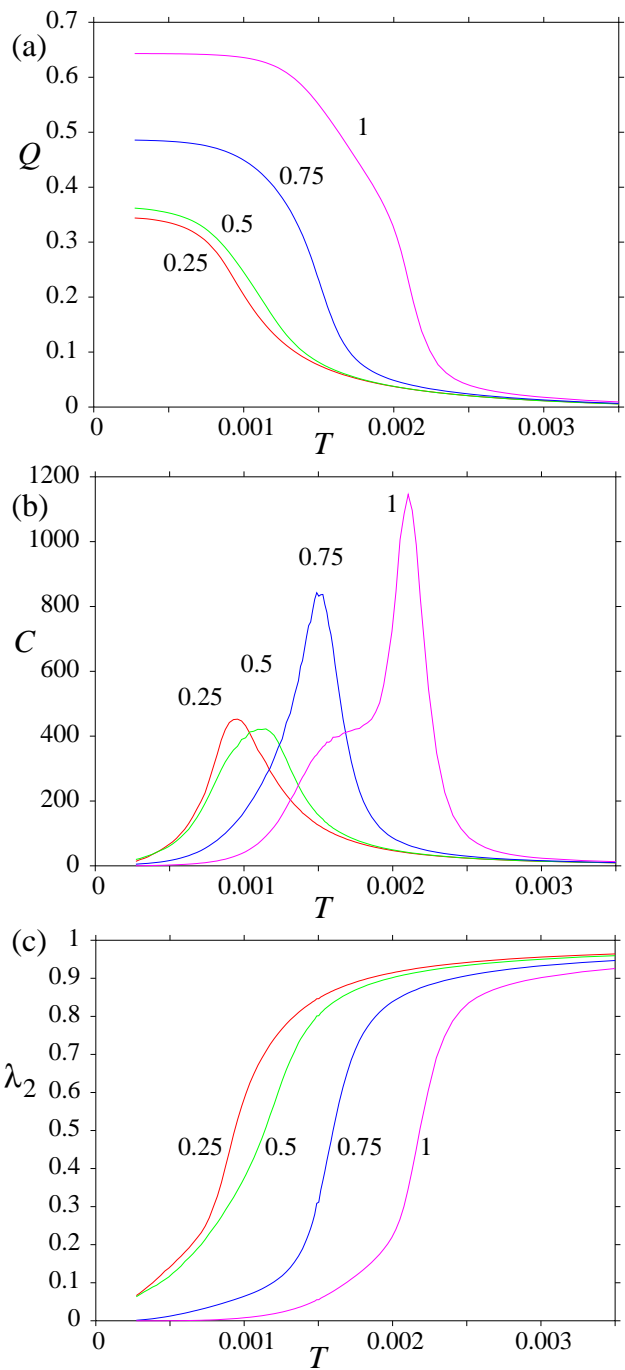}}
\caption{(Colour online) 
Variation of
(a) $Q$,
(b) heat capacity $C$, and 
(c) $\lambda_2$ 
with temperature for the scale-free test networks.
The lines are labelled by the target values of $P_{in}$ used in the network generation.
}
\label{fig:sf}
\end{figure}

Next, we examine randomized versions \cite{Maslov02b,Milo03} of the Apollonian network of Section \ref{sec:apollo} with community structure of varying strength introduced.
Each node is assigned to one of four communities with equal probability, independent of its degree, i.e.~hubs were neither preferentially assigned to the same community nor to different communities.
Edges were then rewired \cite{Maslov02b,Milo03}, while maintaining the degree distribution, and with a predetermined preference for edges to lie within the communities, thus enabling differing values of $P_{in}$ to be generated.
Due to the stochastic nature by which the community structure was introduced, the values of $P_{in}$ of 0.2514, 0.5000, 0.7486 and 0.9126 we obtained did not necessarily precisely match the target values.
Because the networks are scale-free and the nodes are assigned to communities at random, $P_{in}$=0.9126 is the highest value that we could obtain.
These networks have exactly the same degree distribution as the Apollonian network, so it is possible to infer both the effect of the scale-free degree distribution, by comparison with the standard test network of Section \ref{sec:test}, and which features of the Apollonian network are due to its spatial nature and organization.

\begin{figure}
\centerline{\includegraphics[width=8.6cm]{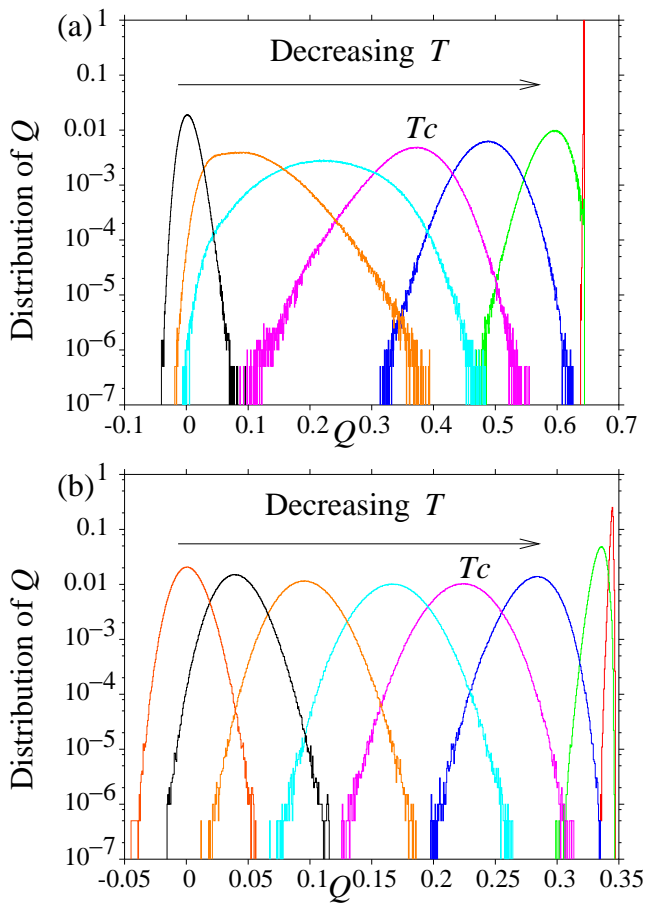}}
\caption{(Colour online)
Distributions of $Q$ at different temperatures for the scale-free test networks with
(a)$P_{in}=0.9126$ and 
(b)$P_{in}=0.2514$.  
}
\label{fig:sfnij}
\end{figure}

In comparison with the standard test networks, the main difference is that the transitions are much broader (Fig.~\ref{fig:sf}), because the greater heterogeneity inherent to scale-free networks leads to a wider range of temperatures at which it becomes favourable for nodes to break away from their assigned communities.
There is less difference between the $P_{in}=0.2514$ and $P_{in}=0.9126$ networks in the position and height of the heat capacity peak than for the standard test networks.

Like the standard test networks, there is still a clear difference between $P_{in} \le 0.5$ and $P_{in} > 0.5$.
For example $Q_{max}$ does not correspond to the input community structure for $P_{in} \le 0.5$.
Furthermore, the frequency matrices at $P_{in} \approx 0.25$ show a very similar temperature evolution.
In particular, there is again evidence that at low temperature there are sets of nodes that can nearly equally well belong to a number of communities.
However, there is less indication of this lack of uniqueness in the low-temperature form for $\lambda_2$ than for the standard networks.

The Apollonian network has a very similar $Q_{max}$ to the test network with $P_{in} \approx 0.75$.
Aside from the hierarchical aspects to the frequency matrices for the Apollonian packing, the properties of the two networks are extremely similar, showing that the degree distribution is the dominant determinant of the behavior.
Even for $P_{in}$ close to one a bimodal distribution of $Q$ values is not seen (Fig.~\ref{fig:sfnij}), unlike the corresponding test network in Section \ref{sec:test}.
The distributions are much broader around $T_c$ when stronger community structure is present.
$Q$ is sensitive to small changes in temperature, leading to a sharp transition, but rather than switching between two distinct phases, this change is continuous.

\begin{figure}
\centerline{\includegraphics[width=8.6cm]{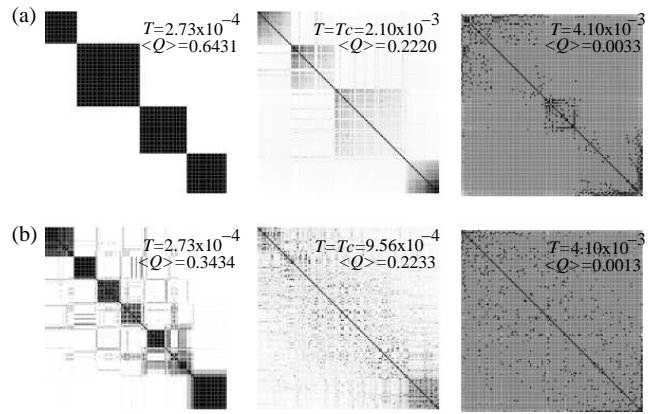}}
\caption{
Frequency matrices at different temperatures for scale-free test networks with (a) $P_{in}=0.9126$, (b) $P_{in}=0.2514$. 
}
\label{fig:fmat-sf}
\end{figure}

\section{Metabolic network}
\label{sec:met}

We now apply this approach to a metabolic network describing pathways between metabolites that has been found to have a scale-free topology \cite{Ma03}.
In this network, nodes represent metabolites and edges connect those that are interconverted by a chemical reaction in \emph{E.\ coli}.
Each chemical reaction can be classified as belonging to a particular metabolic pathway that performs a certain function, e.g.~carbohydrate metabolism.
We first reduced the size of the network from 896 to 304 nodes, by recursively removing all $k=1$ nodes by assuming that they belong to the same community as the one node to which they are connected.
The metabolic network has a very high $Q_{max}$, as noted previously \cite{Ma04,Ma04b,Guimera05,Muff05}, indicating that it is a highly modular network, where the different modules are responsible for different functions.

\begin{figure}
\centerline{\includegraphics[width=8.6cm]{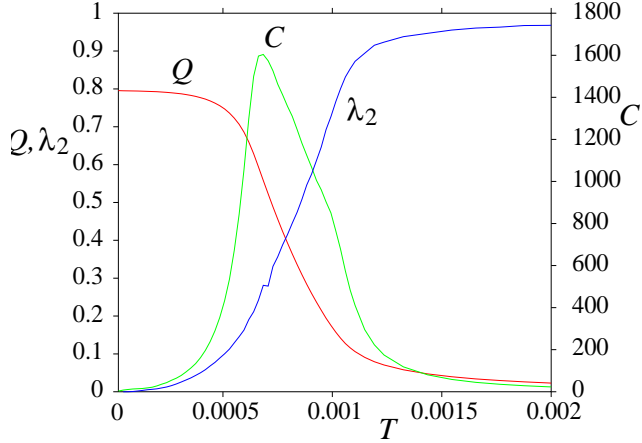}}
\caption{(Colour online) 
Variation of
$Q$,
heat capacity $C$, and 
$\lambda_2$ 
with temperature for the metabolic network.
}
\label{fig:met}
\end{figure}

\begin{figure}
\centerline{\includegraphics[width=8.6cm]{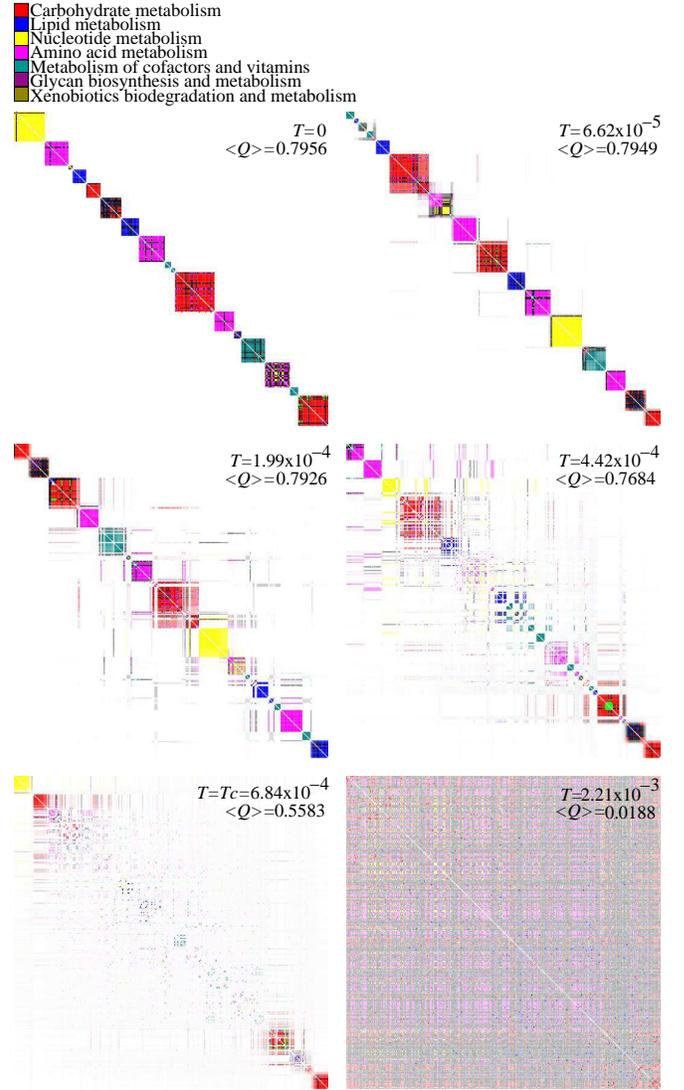}}
\caption{(Colour online)
Frequency matrices at different temperatures for the metabolic network.
If two metabolites are in the same metabolic pathway (see key), the corresponding square is given the colour assigned to that pathway, 
otherwise the square is coloured black.
In the case that the pair of metabolites have more than one pathway in common, the colour of the square is chosen according to that of its neighbours.
In the rare case that the colour remains ambiguous, one of the relevant colours is chosen at random.
}
\label{fig:fmat-met}
\end{figure}

Similar to the other scale-free networks, the transition is fairly broad (Fig.~\ref{fig:met}).
The breadth reflects the broader degree distribution, competing length scales, and heterogeneities in the community structure.
At intermediate temperatures, it is clear that some of the smaller communities break up and merge with other communities, while others remain intact (Fig.~\ref{fig:fmat-met}).
The communities that break up only integrate with a limited number of other nodes, leading to the weak blockiness exemplified in Fig.~\ref{fig:fmat-met}.
In the networks studied so far, the communities have all been fairly similar, whereas in this example some communities are stronger than others, and as such more difficult to break up, thus contributing to the broadness of the transition.

Even though some hierarchy can be seen in the frequency matrix, $\lambda_2$ smoothly tends towards zero at low temperatures, meaning one partition dominates.
This result implies that the uncovered community structure is unique and therefore significant.
Although a few of the low temperature communities involve metabolites 
associated with different pathways,
most communities tend to correspond to a single pathway, 
as illustrated in Fig.~\ref{fig:fmat-met}, 
The stronger communities are subgroups of those corresponding to nucleotide and carbohydrate metabolism, and to a lesser extent amino acid metabolism.

\section{Conclusions}

To assess the significance and nature of the community structure obtained by algorithms that optimize the modularity, we have studied how canonical ensembles of network partitions depend on temperature, where $-Q$ plays the role of energy.
Typically, there is a transition from low entropy, high $Q$ partitions to high entropy, low $Q$ essentially random partitions as the temperature is increased.
The heat capacity provides a useful probe of this transition.
If there is strong community structure, the transition is sharp.
The peak is broader for networks with weaker community structure, as there are more reasonable alternative partitions with intermediate values of $Q$, and so the transition occurs over a broader range of temperature.
If a network has a scale-free or broader degree distribution, the transition also tends to be broader, because the network is more heterogeneous and hubs are likely to make it difficult to separate a network into different modules.

We have introduced an order parameter $\lambda_2$ to quantify the uniqueness of the community structure, 
i.e.~whether there is just a single partition with high $Q$ or a number 
of competing partitions.
$\lambda_2$ is therefore useful as a tool to detect false positives, 
i.e.~partitions whose have high values of $Q$ do not reflect significant 
community structure.
In the case of a lattice, the partition with the highest $Q$ value is highly degenerate and therefore not significant.
Consistent with this, $\lambda_2$ at $T=0$ is non-zero.
For the random networks, $\lambda_2$ is zero at the lowest temperatures because the GM is non-degenerate.
However, below $T_c$ $\lambda_2$ takes significant values (around 0.2--0.3), 
reflecting the many competing partitions that detract from the significance of 
the GM.

Frequency matrices, in particular their temperature dependence,
provide the most information.
For example, it is possible to determine whether the topology is based around a single community structure with small fluctuations, implying that the network is composed of weakly interacting modules, as in the test networks with high $P_{in}$.
Furthermore, more complex modularity features can also be visualised.
For example, in the Apollonian packing, the hierarchical nature is clear form the chequerboard pattern along the diagonal,
and nodes with associations to more than one community are revealed, reflecting overlapping community structures.
The temperature evolution of the frequency matrix for the metabolic network is particularly interesting, as heterogeneities in the community structure are apparent, with the  stronger communities persisting to higher temperature.

In summary, the results in this paper highlight the importance of studying more than simply the `best' partition into communities, and the methods introduced in this paper provide a rigorous approach for doing so.

\begin{acknowledgments}
The authors would like to thank Dr.~Eva Noya for the parallel tempering code, and the Royal Society (JPKD) and the EPSRC for financial support (CPM).
\end{acknowledgments}


\begin{thebibliography}{58}
\expandafter\ifx\csname natexlab\endcsname\relax\def\natexlab#1{#1}\fi
\expandafter\ifx\csname bibnamefont\endcsname\relax
  \def\bibnamefont#1{#1}\fi
\expandafter\ifx\csname bibfnamefont\endcsname\relax
  \def\bibfnamefont#1{#1}\fi
\expandafter\ifx\csname citenamefont\endcsname\relax
  \def\citenamefont#1{#1}\fi
\expandafter\ifx\csname url\endcsname\relax
  \def\url#1{\texttt{#1}}\fi
\expandafter\ifx\csname urlprefix\endcsname\relax\def\urlprefix{URL }\fi
\providecommand{\bibinfo}[2]{#2}
\providecommand{\eprint}[2][]{\url{#2}}

\bibitem[{\citenamefont{Flake et~al.}(2002)\citenamefont{Flake, Lawrence,
  Giles, and Coetzee}}]{Flake02}
\bibinfo{author}{\bibfnamefont{G.~W.} \bibnamefont{Flake}},
  \bibinfo{author}{\bibfnamefont{S.}~\bibnamefont{Lawrence}},
  \bibinfo{author}{\bibfnamefont{C.~L.} \bibnamefont{Giles}}, \bibnamefont{and}
  \bibinfo{author}{\bibfnamefont{F.}~\bibnamefont{Coetzee}},
  \bibinfo{journal}{IEEE Computer} \textbf{\bibinfo{volume}{35}},
  \bibinfo{pages}{66} (\bibinfo{year}{2002}).

\bibitem[{\citenamefont{Girvan and Newman}(2002)}]{Girvan02}
\bibinfo{author}{\bibfnamefont{M.}~\bibnamefont{Girvan}} \bibnamefont{and}
  \bibinfo{author}{\bibfnamefont{M.~E.~J.} \bibnamefont{Newman}},
  \bibinfo{journal}{Proc. Natl. Ac. Sci. U.S.A.} \textbf{\bibinfo{volume}{99}},
  \bibinfo{pages}{7821} (\bibinfo{year}{2002}).

\bibitem[{\citenamefont{Newman and Girvan}(2004)}]{Newman03c}
\bibinfo{author}{\bibfnamefont{M.~E.~J.} \bibnamefont{Newman}}
  \bibnamefont{and} \bibinfo{author}{\bibfnamefont{M.}~\bibnamefont{Girvan}},
  \bibinfo{journal}{Phys. Rev. E} \textbf{\bibinfo{volume}{69}},
  \bibinfo{pages}{026113} (\bibinfo{year}{2004}).

\bibitem[{\citenamefont{Gleiser and Danon}(2003)}]{Gleiser03}
\bibinfo{author}{\bibfnamefont{P.~M.} \bibnamefont{Gleiser}} \bibnamefont{and}
  \bibinfo{author}{\bibfnamefont{L.}~\bibnamefont{Danon}},
  \bibinfo{journal}{Advances in Complex Systems} \textbf{\bibinfo{volume}{6}},
  \bibinfo{pages}{565} (\bibinfo{year}{2003}).

\bibitem[{\citenamefont{Guimer\'a et~al.}(2003)\citenamefont{Guimer\'a, Danon,
  Diaz-Guilera, Giralt, and Arenas}}]{Guimera02}
\bibinfo{author}{\bibfnamefont{R.}~\bibnamefont{Guimer\'a}},
  \bibinfo{author}{\bibfnamefont{L.}~\bibnamefont{Danon}},
  \bibinfo{author}{\bibfnamefont{A.}~\bibnamefont{Diaz-Guilera}},
  \bibinfo{author}{\bibfnamefont{F.}~\bibnamefont{Giralt}}, \bibnamefont{and}
  \bibinfo{author}{\bibfnamefont{A.}~\bibnamefont{Arenas}},
  \bibinfo{journal}{Phys. Rev. E} \textbf{\bibinfo{volume}{68}},
  \bibinfo{pages}{065103(R)} (\bibinfo{year}{2003}).

\bibitem[{\citenamefont{Adamic and Adar}(2003)}]{Adamic03}
\bibinfo{author}{\bibfnamefont{L.~A.} \bibnamefont{Adamic}} \bibnamefont{and}
  \bibinfo{author}{\bibfnamefont{E.}~\bibnamefont{Adar}},
  \bibinfo{journal}{Social Networks} \textbf{\bibinfo{volume}{25}},
  \bibinfo{pages}{211} (\bibinfo{year}{2003}).

\bibitem[{\citenamefont{Hartwell et~al.}(1999)\citenamefont{Hartwell, Hopfield,
  Leibler, and Murray}}]{Hartwell99}
\bibinfo{author}{\bibfnamefont{L.~H.} \bibnamefont{Hartwell}},
  \bibinfo{author}{\bibfnamefont{J.~J.} \bibnamefont{Hopfield}},
  \bibinfo{author}{\bibfnamefont{S.}~\bibnamefont{Leibler}}, \bibnamefont{and}
  \bibinfo{author}{\bibfnamefont{A.~W.} \bibnamefont{Murray}},
  \bibinfo{journal}{Nature} \textbf{\bibinfo{volume}{402}},
  \bibinfo{pages}{C47} (\bibinfo{year}{1999}).

\bibitem[{\citenamefont{Barab\'{a}si and Oltvai}(2004)}]{Barabasi04}
\bibinfo{author}{\bibfnamefont{A.-L.} \bibnamefont{Barab\'{a}si}}
  \bibnamefont{and} \bibinfo{author}{\bibfnamefont{Z.~N.}
  \bibnamefont{Oltvai}}, \bibinfo{journal}{Nature Reviews:Genetics}
  \textbf{\bibinfo{volume}{5}}, \bibinfo{pages}{101} (\bibinfo{year}{2004}).

\bibitem[{\citenamefont{Ma et~al.}(2004{\natexlab{a}})\citenamefont{Ma, Buer,
  and Zeng}}]{Ma04}
\bibinfo{author}{\bibfnamefont{H.~W.} \bibnamefont{Ma}},
  \bibinfo{author}{\bibfnamefont{J.}~\bibnamefont{Buer}}, \bibnamefont{and}
  \bibinfo{author}{\bibfnamefont{A.-P.} \bibnamefont{Zeng}},
  \bibinfo{journal}{BMC Bioinformatics} \textbf{\bibinfo{volume}{5}},
  \bibinfo{pages}{199} (\bibinfo{year}{2004}{\natexlab{a}}).

\bibitem[{\citenamefont{Guimer\'a and Amaral}(2005)}]{Guimera05}
\bibinfo{author}{\bibfnamefont{R.}~\bibnamefont{Guimer\'a}} \bibnamefont{and}
  \bibinfo{author}{\bibfnamefont{L.~A.~N.} \bibnamefont{Amaral}},
  \bibinfo{journal}{Nature} \textbf{\bibinfo{volume}{433}},
  \bibinfo{pages}{895} (\bibinfo{year}{2005}).

\bibitem[{\citenamefont{Milo et~al.}(2002)\citenamefont{Milo, Shen-Orr,
  Itzkovitz, Kashtan, Chklovskii, and Alon}}]{Milo02}
\bibinfo{author}{\bibfnamefont{R.}~\bibnamefont{Milo}},
  \bibinfo{author}{\bibfnamefont{S.}~\bibnamefont{Shen-Orr}},
  \bibinfo{author}{\bibfnamefont{S.}~\bibnamefont{Itzkovitz}},
  \bibinfo{author}{\bibfnamefont{N.}~\bibnamefont{Kashtan}},
  \bibinfo{author}{\bibfnamefont{D.}~\bibnamefont{Chklovskii}},
  \bibnamefont{and} \bibinfo{author}{\bibfnamefont{U.}~\bibnamefont{Alon}},
  \bibinfo{journal}{Science} \textbf{\bibinfo{volume}{298}},
  \bibinfo{pages}{824} (\bibinfo{year}{2002}).

\bibitem[{\citenamefont{Wilkinson and
  Huberman}(2004{\natexlab{a}})}]{Wilkinson04}
\bibinfo{author}{\bibfnamefont{D.}~\bibnamefont{Wilkinson}} \bibnamefont{and}
  \bibinfo{author}{\bibfnamefont{B.~A.} \bibnamefont{Huberman}},
  \bibinfo{journal}{Proc. Natl. Acad. Sci. U.S.A.}
  \textbf{\bibinfo{volume}{101}}, \bibinfo{pages}{5241}
  (\bibinfo{year}{2004}{\natexlab{a}}).

\bibitem[{\citenamefont{Palla et~al.}(2005)\citenamefont{Palla, Der\'enyi,
  Farkas, and Vicsek}}]{Palla05}
\bibinfo{author}{\bibfnamefont{G.}~\bibnamefont{Palla}},
  \bibinfo{author}{\bibfnamefont{I.}~\bibnamefont{Der\'enyi}},
  \bibinfo{author}{\bibfnamefont{I.}~\bibnamefont{Farkas}}, \bibnamefont{and}
  \bibinfo{author}{\bibfnamefont{T.}~\bibnamefont{Vicsek}},
  \bibinfo{journal}{Nature} \textbf{\bibinfo{volume}{435}},
  \bibinfo{pages}{814} (\bibinfo{year}{2005}).

\bibitem[{\citenamefont{Huss and Holme}()}]{Huss06}
\bibinfo{author}{\bibfnamefont{M.}~\bibnamefont{Huss}} \bibnamefont{and}
  \bibinfo{author}{\bibfnamefont{P.}~\bibnamefont{Holme}},
  \bibinfo{journal}{q-bio.MN/0603038}.

\bibitem[{\citenamefont{Granovetter}(1973)}]{Granovetter73}
\bibinfo{author}{\bibfnamefont{M.~S.} \bibnamefont{Granovetter}},
  \bibinfo{journal}{Am. J. Soc.} \textbf{\bibinfo{volume}{78}},
  \bibinfo{pages}{1360} (\bibinfo{year}{1973}).

\bibitem[{\citenamefont{Tyler et~al.}(2003)\citenamefont{Tyler, Wilkinson, and
  Huberman}}]{Tyler03book}
\bibinfo{author}{\bibfnamefont{J.~R.} \bibnamefont{Tyler}},
  \bibinfo{author}{\bibfnamefont{D.~M.} \bibnamefont{Wilkinson}},
  \bibnamefont{and} \bibinfo{author}{\bibfnamefont{B.~A.}
  \bibnamefont{Huberman}}, in \emph{\bibinfo{booktitle}{Communities and
  technologies}}, edited by
  \bibinfo{editor}{\bibfnamefont{B.}~\bibnamefont{Kluwer}}
  (\bibinfo{publisher}{Deventer}, \bibinfo{address}{The Netherlands},
  \bibinfo{year}{2003}), pp. \bibinfo{pages}{81--96}.

\bibitem[{\citenamefont{Xie et~al.}(2006)\citenamefont{Xie, Yan, and
  Maslov}}]{Xie04}
\bibinfo{author}{\bibfnamefont{H.}~\bibnamefont{Xie}},
  \bibinfo{author}{\bibfnamefont{K.-K.} \bibnamefont{Yan}}, \bibnamefont{and}
  \bibinfo{author}{\bibfnamefont{S.}~\bibnamefont{Maslov}}, in
  \emph{\bibinfo{booktitle}{Proceedings of the Winter School ``Dynamics of
  Complex Interconnected Systems: Networks and Bioprocesses, Geilo 2005}}
  (\bibinfo{publisher}{Kluwer Academic Publishers},
  \bibinfo{address}{Dordrecht}, \bibinfo{year}{2006}).

\bibitem[{\citenamefont{Bettencourt}()}]{Bettencourt03}
\bibinfo{author}{\bibfnamefont{L.~M.~A.} \bibnamefont{Bettencourt}},
  \bibinfo{journal}{cond-mat/0304321}.

\bibitem[{\citenamefont{Eriksen et~al.}(2003)\citenamefont{Eriksen, Simonsen,
  Maslov, and Sneppen}}]{Eriksen03}
\bibinfo{author}{\bibfnamefont{K.~A.} \bibnamefont{Eriksen}},
  \bibinfo{author}{\bibfnamefont{I.}~\bibnamefont{Simonsen}},
  \bibinfo{author}{\bibfnamefont{S.}~\bibnamefont{Maslov}}, \bibnamefont{and}
  \bibinfo{author}{\bibfnamefont{K.}~\bibnamefont{Sneppen}},
  \bibinfo{journal}{Phys. Rev. Lett.} \textbf{\bibinfo{volume}{90}},
  \bibinfo{pages}{148701} (\bibinfo{year}{2003}).

\bibitem[{\citenamefont{Pons and Latapy}(2005)}]{Latapy04}
\bibinfo{author}{\bibfnamefont{P.}~\bibnamefont{Pons}} \bibnamefont{and}
  \bibinfo{author}{\bibfnamefont{M.}~\bibnamefont{Latapy}},
  \bibinfo{journal}{Lect. Notes in Comp. Sci.} \textbf{\bibinfo{volume}{3733}},
  \bibinfo{pages}{284} (\bibinfo{year}{2005}).

\bibitem[{\citenamefont{Simonsen et~al.}(2004)\citenamefont{Simonsen, Eriksen,
  Maslov, and Sneppen}}]{Simonsen04}
\bibinfo{author}{\bibfnamefont{I.}~\bibnamefont{Simonsen}},
  \bibinfo{author}{\bibfnamefont{K.~A.} \bibnamefont{Eriksen}},
  \bibinfo{author}{\bibfnamefont{S.}~\bibnamefont{Maslov}}, \bibnamefont{and}
  \bibinfo{author}{\bibfnamefont{K.}~\bibnamefont{Sneppen}},
  \bibinfo{journal}{Physica A} \textbf{\bibinfo{volume}{336}},
  \bibinfo{pages}{163} (\bibinfo{year}{2004}).

\bibitem[{\citenamefont{Huang et~al.}(2006)\citenamefont{Huang, Park, and
  Lai}}]{Huang06}
\bibinfo{author}{\bibfnamefont{L.}~\bibnamefont{Huang}},
  \bibinfo{author}{\bibfnamefont{K.}~\bibnamefont{Park}}, \bibnamefont{and}
  \bibinfo{author}{\bibfnamefont{Y.-C.} \bibnamefont{Lai}},
  \bibinfo{journal}{Phys. Rev. E} \textbf{\bibinfo{volume}{73}},
  \bibinfo{pages}{035103(R)} (\bibinfo{year}{2006}).

\bibitem[{\citenamefont{Watts et~al.}(2002)\citenamefont{Watts, Dodds, and
  Newman}}]{Watts02}
\bibinfo{author}{\bibfnamefont{D.~J.} \bibnamefont{Watts}},
  \bibinfo{author}{\bibfnamefont{P.~S.} \bibnamefont{Dodds}}, \bibnamefont{and}
  \bibinfo{author}{\bibfnamefont{M.~E.~J.} \bibnamefont{Newman}},
  \bibinfo{journal}{Science} \textbf{\bibinfo{volume}{296}},
  \bibinfo{pages}{1302} (\bibinfo{year}{2002}).

\bibitem[{\citenamefont{Motter et~al.}(2003)\citenamefont{Motter, Nishikawa,
  and Lai}}]{Motter03}
\bibinfo{author}{\bibfnamefont{A.~E.} \bibnamefont{Motter}},
  \bibinfo{author}{\bibfnamefont{T.}~\bibnamefont{Nishikawa}},
  \bibnamefont{and} \bibinfo{author}{\bibfnamefont{Y.-C.} \bibnamefont{Lai}},
  \bibinfo{journal}{Phys. Rev. E} \textbf{\bibinfo{volume}{68}},
  \bibinfo{pages}{036105} (\bibinfo{year}{2003}).

\bibitem[{\citenamefont{Song et~al.}(2005)\citenamefont{Song, Havlin, and
  Makse}}]{Song05}
\bibinfo{author}{\bibfnamefont{C.}~\bibnamefont{Song}},
  \bibinfo{author}{\bibfnamefont{S.}~\bibnamefont{Havlin}}, \bibnamefont{and}
  \bibinfo{author}{\bibfnamefont{H.~A.} \bibnamefont{Makse}},
  \bibinfo{journal}{Nature} \textbf{\bibinfo{volume}{433}},
  \bibinfo{pages}{392} (\bibinfo{year}{2005}).

\bibitem[{\citenamefont{Itzkovitz et~al.}(2005)\citenamefont{Itzkovitz, Levitt,
  Kashtan, Milo, Itzkovitz, and Alon}}]{Itzkovitz05}
\bibinfo{author}{\bibfnamefont{S.}~\bibnamefont{Itzkovitz}},
  \bibinfo{author}{\bibfnamefont{R.}~\bibnamefont{Levitt}},
  \bibinfo{author}{\bibfnamefont{N.}~\bibnamefont{Kashtan}},
  \bibinfo{author}{\bibfnamefont{R.}~\bibnamefont{Milo}},
  \bibinfo{author}{\bibfnamefont{M.}~\bibnamefont{Itzkovitz}},
  \bibnamefont{and} \bibinfo{author}{\bibfnamefont{U.}~\bibnamefont{Alon}},
  \bibinfo{journal}{Phys. Rev. E} \textbf{\bibinfo{volume}{71}},
  \bibinfo{pages}{016127} (\bibinfo{year}{2005}).

\bibitem[{\citenamefont{Newman}(2004{\natexlab{a}})}]{Newman04}
\bibinfo{author}{\bibfnamefont{M.~E.~J.} \bibnamefont{Newman}},
  \bibinfo{journal}{Eur. Phys. J. B.} \textbf{\bibinfo{volume}{38}},
  \bibinfo{pages}{321} (\bibinfo{year}{2004}{\natexlab{a}}).

\bibitem[{\citenamefont{Danon et~al.}(2005)\citenamefont{Danon, Diaz-Guilera,
  Duch, and Arenas}}]{Danon05}
\bibinfo{author}{\bibfnamefont{L.}~\bibnamefont{Danon}},
  \bibinfo{author}{\bibfnamefont{A.}~\bibnamefont{Diaz-Guilera}},
  \bibinfo{author}{\bibfnamefont{J.}~\bibnamefont{Duch}}, \bibnamefont{and}
  \bibinfo{author}{\bibfnamefont{A.}~\bibnamefont{Arenas}},
  \bibinfo{journal}{J. Stat. Mech.: Theory and Experiment},
  \bibinfo{pages}{P09008} (\bibinfo{year}{2005}).

\bibitem[{\citenamefont{Newman}(2004{\natexlab{b}})}]{Newman03d}
\bibinfo{author}{\bibfnamefont{M.~E.~J.} \bibnamefont{Newman}},
  \bibinfo{journal}{Phys. Rev. E} \textbf{\bibinfo{volume}{69}},
  \bibinfo{pages}{066133} (\bibinfo{year}{2004}{\natexlab{b}}).

\bibitem[{\citenamefont{Guimer\'a et~al.}(2004)\citenamefont{Guimer\'a,
  Sales-Pardo, and Amaral}}]{Guimera04}
\bibinfo{author}{\bibfnamefont{R.}~\bibnamefont{Guimer\'a}},
  \bibinfo{author}{\bibfnamefont{M.}~\bibnamefont{Sales-Pardo}},
  \bibnamefont{and} \bibinfo{author}{\bibfnamefont{L.~A.~N.}
  \bibnamefont{Amaral}}, \bibinfo{journal}{Phys. Rev. E}
  \textbf{\bibinfo{volume}{70}}, \bibinfo{pages}{025101(R)}
  (\bibinfo{year}{2004}).

\bibitem[{\citenamefont{Massen and Doye}(2005)}]{Massen05}
\bibinfo{author}{\bibfnamefont{C.~P.} \bibnamefont{Massen}} \bibnamefont{and}
  \bibinfo{author}{\bibfnamefont{J.~P.~K.} \bibnamefont{Doye}},
  \bibinfo{journal}{Phys. Rev. E} \textbf{\bibinfo{volume}{71}},
  \bibinfo{pages}{046101} (\bibinfo{year}{2005}).

\bibitem[{\citenamefont{Reichardt and Bornholdt}()}]{Reichardt06b}
\bibinfo{author}{\bibfnamefont{J.}~\bibnamefont{Reichardt}} \bibnamefont{and}
  \bibinfo{author}{\bibfnamefont{S.}~\bibnamefont{Bornholdt}},
  \bibinfo{journal}{cond-mat/0606220}.

\bibitem[{\citenamefont{Wilkinson and
  Huberman}(2004{\natexlab{b}})}]{Wilkinson02}
\bibinfo{author}{\bibfnamefont{D.~M.} \bibnamefont{Wilkinson}}
  \bibnamefont{and} \bibinfo{author}{\bibfnamefont{B.~A.}
  \bibnamefont{Huberman}}, \bibinfo{journal}{Proc. Natl. Acad. Sci. U.S.A.}
  \textbf{\bibinfo{volume}{101}}, \bibinfo{pages}{5241}
  (\bibinfo{year}{2004}{\natexlab{b}}).

\bibitem[{\citenamefont{Reichardt and Bornholdt}(2004)}]{Reichardt04}
\bibinfo{author}{\bibfnamefont{J.}~\bibnamefont{Reichardt}} \bibnamefont{and}
  \bibinfo{author}{\bibfnamefont{S.}~\bibnamefont{Bornholdt}},
  \bibinfo{journal}{Phys. Rev. Lett.} \textbf{\bibinfo{volume}{93}},
  \bibinfo{pages}{218701} (\bibinfo{year}{2004}).

\bibitem[{\citenamefont{Der\'enyi et~al.}(2005)\citenamefont{Der\'enyi, Palla,
  and Vicsek}}]{Derenyi05}
\bibinfo{author}{\bibfnamefont{I.}~\bibnamefont{Der\'enyi}},
  \bibinfo{author}{\bibfnamefont{G.}~\bibnamefont{Palla}}, \bibnamefont{and}
  \bibinfo{author}{\bibfnamefont{T.}~\bibnamefont{Vicsek}},
  \bibinfo{journal}{Phys. Rev. Lett.} \textbf{\bibinfo{volume}{94}},
  \bibinfo{pages}{160202} (\bibinfo{year}{2005}).

\bibitem[{\citenamefont{Gfeller et~al.}(2005)\citenamefont{Gfeller, Chappelier,
  and De~Los~Rios}}]{Gfeller05}
\bibinfo{author}{\bibfnamefont{D.}~\bibnamefont{Gfeller}},
  \bibinfo{author}{\bibfnamefont{J.-C.} \bibnamefont{Chappelier}},
  \bibnamefont{and}
  \bibinfo{author}{\bibfnamefont{P.}~\bibnamefont{De~Los~Rios}},
  \bibinfo{journal}{Phys. Rev. E} \textbf{\bibinfo{volume}{72}},
  \bibinfo{pages}{056135} (\bibinfo{year}{2005}).

\bibitem[{\citenamefont{Ravasz et~al.}(2002)\citenamefont{Ravasz, Somera,
  Mongru, Oltvai, and Barab\'asi}}]{Ravasz02}
\bibinfo{author}{\bibfnamefont{E.}~\bibnamefont{Ravasz}},
  \bibinfo{author}{\bibfnamefont{A.~L.} \bibnamefont{Somera}},
  \bibinfo{author}{\bibfnamefont{D.~A.} \bibnamefont{Mongru}},
  \bibinfo{author}{\bibfnamefont{Z.~N.} \bibnamefont{Oltvai}},
  \bibnamefont{and} \bibinfo{author}{\bibfnamefont{A.-L.}
  \bibnamefont{Barab\'asi}}, \bibinfo{journal}{Science}
  \textbf{\bibinfo{volume}{297}}, \bibinfo{pages}{1551} (\bibinfo{year}{2002}).

\bibitem[{\citenamefont{Holme and Huss}(2003)}]{Holme03}
\bibinfo{author}{\bibfnamefont{P.}~\bibnamefont{Holme}} \bibnamefont{and}
  \bibinfo{author}{\bibfnamefont{M.}~\bibnamefont{Huss}}, \bibinfo{journal}{3rd
  Workshop on Computation of Biochemical Pathways and Genetic Networks} pp.
  \bibinfo{pages}{3--9} (\bibinfo{year}{2003}).

\bibitem[{\citenamefont{Ravasz and Barab\'asi}(2003)}]{Ravasz03}
\bibinfo{author}{\bibfnamefont{E.}~\bibnamefont{Ravasz}} \bibnamefont{and}
  \bibinfo{author}{\bibfnamefont{A.-L.} \bibnamefont{Barab\'asi}},
  \bibinfo{journal}{Phys. Rev. E} \textbf{\bibinfo{volume}{67}},
  \bibinfo{pages}{026112} (\bibinfo{year}{2003}).

\bibitem[{\citenamefont{Zhou}(2003)}]{Zhou03}
\bibinfo{author}{\bibfnamefont{H.}~\bibnamefont{Zhou}}, \bibinfo{journal}{Phys.
  Rev. E} \textbf{\bibinfo{volume}{67}}, \bibinfo{pages}{061901}
  (\bibinfo{year}{2003}).

\bibitem[{\citenamefont{Muff et~al.}(2005)\citenamefont{Muff, Rao, and
  Caflisch}}]{Muff05}
\bibinfo{author}{\bibfnamefont{S.}~\bibnamefont{Muff}},
  \bibinfo{author}{\bibfnamefont{F.}~\bibnamefont{Rao}}, \bibnamefont{and}
  \bibinfo{author}{\bibfnamefont{A.}~\bibnamefont{Caflisch}},
  \bibinfo{journal}{Phys. Rev. E} \textbf{\bibinfo{volume}{72}},
  \bibinfo{pages}{056107} (\bibinfo{year}{2005}).

\bibitem[{\citenamefont{Duch and Arenas}(2005)}]{Duch05}
\bibinfo{author}{\bibfnamefont{J.}~\bibnamefont{Duch}} \bibnamefont{and}
  \bibinfo{author}{\bibfnamefont{A.}~\bibnamefont{Arenas}},
  \bibinfo{journal}{Phys. Rev. E} \textbf{\bibinfo{volume}{72}},
  \bibinfo{pages}{027104} (\bibinfo{year}{2005}).

\bibitem[{\citenamefont{Reichardt and Bornholdt}(2006)}]{Reichardt06}
\bibinfo{author}{\bibfnamefont{J.}~\bibnamefont{Reichardt}} \bibnamefont{and}
  \bibinfo{author}{\bibfnamefont{S.}~\bibnamefont{Bornholdt}},
  \bibinfo{journal}{Phys. Rev. E} \textbf{\bibinfo{volume}{74}},
  \bibinfo{pages}{016110} (\bibinfo{year}{2006}).

\bibitem[{\citenamefont{Medus et~al.}(2005)\citenamefont{Medus, Acuna, and
  Dorso}}]{Medus05}
\bibinfo{author}{\bibfnamefont{A.}~\bibnamefont{Medus}},
  \bibinfo{author}{\bibfnamefont{G.}~\bibnamefont{Acuna}}, \bibnamefont{and}
  \bibinfo{author}{\bibfnamefont{C.~O.} \bibnamefont{Dorso}},
  \bibinfo{journal}{Physica A} \textbf{\bibinfo{volume}{358}},
  \bibinfo{pages}{593} (\bibinfo{year}{2005}).

\bibitem[{\citenamefont{Doye and Massen}(2005)}]{Doye05}
\bibinfo{author}{\bibfnamefont{J.~P.~K.} \bibnamefont{Doye}} \bibnamefont{and}
  \bibinfo{author}{\bibfnamefont{C.~P.} \bibnamefont{Massen}},
  \bibinfo{journal}{Phys. Rev. E} \textbf{\bibinfo{volume}{71}},
  \bibinfo{pages}{016128} (\bibinfo{year}{2005}).

\bibitem[{\citenamefont{Andrade~Jr. et~al.}(2005)\citenamefont{Andrade~Jr.,
  Herrmann, Andrade, and da~Silva}}]{Andrade05}
\bibinfo{author}{\bibfnamefont{J.~S.} \bibnamefont{Andrade~Jr.}},
  \bibinfo{author}{\bibfnamefont{H.~J.} \bibnamefont{Herrmann}},
  \bibinfo{author}{\bibfnamefont{R.~F.~S.} \bibnamefont{Andrade}},
  \bibnamefont{and} \bibinfo{author}{\bibfnamefont{L.~R.}
  \bibnamefont{da~Silva}}, \bibinfo{journal}{Phys. Rev. Lett.}
  \textbf{\bibinfo{volume}{94}}, \bibinfo{pages}{018702}
  (\bibinfo{year}{2005}).

\bibitem[{\citenamefont{Ma and Zeng}(2003)}]{Ma03}
\bibinfo{author}{\bibfnamefont{H.}~\bibnamefont{Ma}} \bibnamefont{and}
  \bibinfo{author}{\bibfnamefont{A.-P.} \bibnamefont{Zeng}},
  \bibinfo{journal}{Bioinformatics} \textbf{\bibinfo{volume}{19}},
  \bibinfo{pages}{270} (\bibinfo{year}{2003}).

\bibitem[{\citenamefont{Ma et~al.}(2004{\natexlab{b}})\citenamefont{Ma, Zhao,
  Yuan, and Zeng}}]{Ma04b}
\bibinfo{author}{\bibfnamefont{H.~W.} \bibnamefont{Ma}},
  \bibinfo{author}{\bibfnamefont{X.~M.} \bibnamefont{Zhao}},
  \bibinfo{author}{\bibfnamefont{Y.~J.} \bibnamefont{Yuan}}, \bibnamefont{and}
  \bibinfo{author}{\bibfnamefont{A.-P.} \bibnamefont{Zeng}},
  \bibinfo{journal}{Bioinformatics} \textbf{\bibinfo{volume}{20}},
  \bibinfo{pages}{1870} (\bibinfo{year}{2004}{\natexlab{b}}).

\bibitem[{\citenamefont{Holme et~al.}(2003)\citenamefont{Holme, Huss, and
  Jeong}}]{Holme03b}
\bibinfo{author}{\bibfnamefont{P.}~\bibnamefont{Holme}},
  \bibinfo{author}{\bibfnamefont{M.}~\bibnamefont{Huss}}, \bibnamefont{and}
  \bibinfo{author}{\bibfnamefont{H.}~\bibnamefont{Jeong}},
  \bibinfo{journal}{Bioinformatics} \textbf{\bibinfo{volume}{19}},
  \bibinfo{pages}{532} (\bibinfo{year}{2003}).

\bibitem[{\citenamefont{Earl and Deem}(2005)}]{Earl05}
\bibinfo{author}{\bibfnamefont{D.~J.} \bibnamefont{Earl}} \bibnamefont{and}
  \bibinfo{author}{\bibfnamefont{M.~W.} \bibnamefont{Deem}},
  \bibinfo{journal}{Phys. Chem. Chem. Phys.} \textbf{\bibinfo{volume}{7}},
  \bibinfo{pages}{3910} (\bibinfo{year}{2005}).

\bibitem[{\citenamefont{Brandes et~al.}()\citenamefont{Brandes, Delling,
  Gaertler, Hoefer, Nikoloski, and Wagner}}]{Brandes06}
\bibinfo{author}{\bibfnamefont{U.}~\bibnamefont{Brandes}},
  \bibinfo{author}{\bibfnamefont{D.}~\bibnamefont{Delling}},
  \bibinfo{author}{\bibfnamefont{M.}~\bibnamefont{Gaertler}},
  \bibinfo{author}{\bibfnamefont{M.}~\bibnamefont{Hoefer}},
  \bibinfo{author}{\bibfnamefont{Z.}~\bibnamefont{Nikoloski}},
  \bibnamefont{and} \bibinfo{author}{\bibfnamefont{D.}~\bibnamefont{Wagner}},
  \bibinfo{journal}{physics/0608255}.

\bibitem[{\citenamefont{Fiedler}(1973)}]{Fiedler73}
\bibinfo{author}{\bibfnamefont{M.}~\bibnamefont{Fiedler}},
  \bibinfo{journal}{Czech. Math. J.} \textbf{\bibinfo{volume}{23}},
  \bibinfo{pages}{298} (\bibinfo{year}{1973}).

\bibitem[{\citenamefont{Capocci et~al.}(2005)\citenamefont{Capocci, Servedio,
  Caldarelli, and Colaiori}}]{Capocci04}
\bibinfo{author}{\bibfnamefont{A.}~\bibnamefont{Capocci}},
  \bibinfo{author}{\bibfnamefont{V.~D.~P.} \bibnamefont{Servedio}},
  \bibinfo{author}{\bibfnamefont{G.}~\bibnamefont{Caldarelli}},
  \bibnamefont{and} \bibinfo{author}{\bibfnamefont{F.}~\bibnamefont{Colaiori}},
  \bibinfo{journal}{Physica A} \textbf{\bibinfo{volume}{352}},
  \bibinfo{pages}{669} (\bibinfo{year}{2005}).

\bibitem[{\citenamefont{Newman}(2006{\natexlab{a}})}]{Newman06}
\bibinfo{author}{\bibfnamefont{M.~E.~J.} \bibnamefont{Newman}},
  \bibinfo{journal}{Proc. Nat. Acad. Sci. U.S.A.}
  \textbf{\bibinfo{volume}{103}}, \bibinfo{pages}{8577}
  (\bibinfo{year}{2006}{\natexlab{a}}).

\bibitem[{\citenamefont{Newman}(2006{\natexlab{b}})}]{Newman06b}
\bibinfo{author}{\bibfnamefont{M.~E.~J.} \bibnamefont{Newman}},
  \bibinfo{journal}{Phys. Rev. E} \textbf{\bibinfo{volume}{74}},
  \bibinfo{pages}{036104} (\bibinfo{year}{2006}{\natexlab{b}}).

\bibitem[{\citenamefont{Mandelbrot}(1977)}]{Mandelbrot}
\bibinfo{author}{\bibfnamefont{B.~B.} \bibnamefont{Mandelbrot}},
  \emph{\bibinfo{title}{The Fractal Geometry of Nature}}
  (\bibinfo{publisher}{W. H. Freeman and Company}, \bibinfo{address}{New York},
  \bibinfo{year}{1977}).

\bibitem[{\citenamefont{Maslov and Sneppen}(2002)}]{Maslov02b}
\bibinfo{author}{\bibfnamefont{S.}~\bibnamefont{Maslov}} \bibnamefont{and}
  \bibinfo{author}{\bibfnamefont{K.}~\bibnamefont{Sneppen}},
  \bibinfo{journal}{Science} \textbf{\bibinfo{volume}{296}},
  \bibinfo{pages}{910} (\bibinfo{year}{2002}).

\bibitem[{\citenamefont{Milo et~al.}()\citenamefont{Milo, Kashtan, Itzkovitz,
  Newman, and Alon}}]{Milo03}
\bibinfo{author}{\bibfnamefont{R.}~\bibnamefont{Milo}},
  \bibinfo{author}{\bibfnamefont{N.}~\bibnamefont{Kashtan}},
  \bibinfo{author}{\bibfnamefont{S.}~\bibnamefont{Itzkovitz}},
  \bibinfo{author}{\bibfnamefont{M.~E.~J.} \bibnamefont{Newman}},
  \bibnamefont{and} \bibinfo{author}{\bibfnamefont{U.}~\bibnamefont{Alon}},
  \bibinfo{journal}{cond-mat/0312028}.

\end{thebibliography}
\end{document}